\documentclass[sigconf]{acmart}
\AtBeginDocument{%
  }

\setcopyright{acmlicensed}
\copyrightyear{2018}
\acmYear{2018}
\acmDOI{XXXXXXX.XXXXXXX}
\acmConference[Conference acronym 'XX]{Make sure to enter the correct
  conference title from your rights confirmation email}{June 03--05,
  2018}{Woodstock, NY}
\acmISBN{978-1-4503-XXXX-X/2018/06}

\usepackage{listings}
\usepackage{algorithm}
\usepackage{algpseudocode}
\begin{document}

\title{Automated Type Annotation in Python Using Large Language Models
}

\author{Varun Bharti}
\authornote{Both authors contributed equally to this research}
\affiliation{%
  \institution{IIIT Delhi}
  \city{Delhi}
  \country{India}}
\email{varun22562@iiitd.ac.in}

\author{Shashwat Jha}
\authornotemark[1]
\affiliation{%
  \institution{IIIT Delhi}
  \city{Delhi}
  \country{India}}
\email{shashwat22472@iiitd.ac.in}
\author{Dhruv Kumar}
\affiliation{%
  \institution{BITS Pilani}
  \city{Pilani}
  \country{India}}
\email{dhruv.kumar@pilani.bits-pilani.ac.in}
\author{Pankaj Jalote}
\affiliation{%
  \institution{IIIT Delhi}
  \city{Delhi}
  \country{India}}
\email{jalote@iiitd.ac.in}

%%

\begin{abstract}
Type annotations in Python enhance maintainability and error detection. However, generating these annotations manually is error-prone and requires extra effort. Traditional automation approaches like static analysis, machine learning, and deep learning struggle with limited type vocabularies, behavioral over-approximation, and reliance on large labeled datasets.

In this work, we explore the use of LLMs for generating type annotations in Python. We develop a generate–check–repair pipeline: the LLM proposes annotations guided by a Concrete Syntax Tree representation, a static type checker (\texttt{Mypy}) verifies them, and any errors are fed back for iterative refinement. We evaluate four LLM variants: GPT-4o-Mini, GPT-4.1-mini (general-purpose), and O3-Mini, O4-Mini (reasoning-optimized), on 6000 code snippets from the ManyTypes4Py benchmark. We first measure the proportion of code snippets annotated by LLMs for which MyPy reported no errors (i.e., consistent results): GPT‑4o-Mini achieved consistency on 65.9\% of cases (34.1\% inconsistent), while GPT‑4.1‑mini, O3-Mini, and O4-Mini each reached approximately 88.6\% consistency (around 11.4\% failures). To measure annotation quality, we then compute exact‑match and base‑type match accuracies over all 6000 snippets: GPT-4.1-mini and O3-Mini perform the best, achieving up to 70.5\% exact-match and 79.1\% base-type accuracy, requiring under one repair iteration on average. 

Our results demonstrate that general-purpose and reasoning-optimized LLMs, without any task-specific fine-tuning or additional training can be effective in generating consistent type annotations. They perform competitively with traditional deep learning techniques which require large labeled dataset for training. While our work focuses on Python, the pipeline can be extended to other optionally typed imperative languages like Ruby.
\end{abstract}

\begin{CCSXML}
<ccs2012>
   <concept>
       <concept_id>10011007.10011006.10011008.10011024.10003202</concept_id>
       <concept_desc>Software and its engineering~Abstract data types</concept_desc>
       <concept_significance>500</concept_significance>
       </concept>
   <concept>
       <concept_id>10011007.10011006.10011008.10011024.10011028</concept_id>
       <concept_desc>Software and its engineering~Data types and structures</concept_desc>
       <concept_significance>500</concept_significance>
       </concept>
   <concept>
       <concept_id>10011007.10011074.10011111.10011696</concept_id>
       <concept_desc>Software and its engineering~Maintaining software</concept_desc>
       <concept_significance>500</concept_significance>
       </concept>
   <concept>
       <concept_id>10010147.10010178.10010179</concept_id>
       <concept_desc>Computing methodologies~Natural language processing</concept_desc>
       <concept_significance>500</concept_significance>
       </concept>
 </ccs2012>
\end{CCSXML}

\ccsdesc[500]{Software and its engineering~Abstract data types}
\ccsdesc[500]{Software and its engineering~Data types and structures}
\ccsdesc[500]{Software and its engineering~Maintaining software}
\ccsdesc[500]{Computing methodologies~Natural language processing}

\keywords{ Type Annotations, Large Language Models, Static Type Checking, Mypy, Python }

\maketitle

\begin{figure*}[t]
  \centering
  \includegraphics[width=\textwidth]{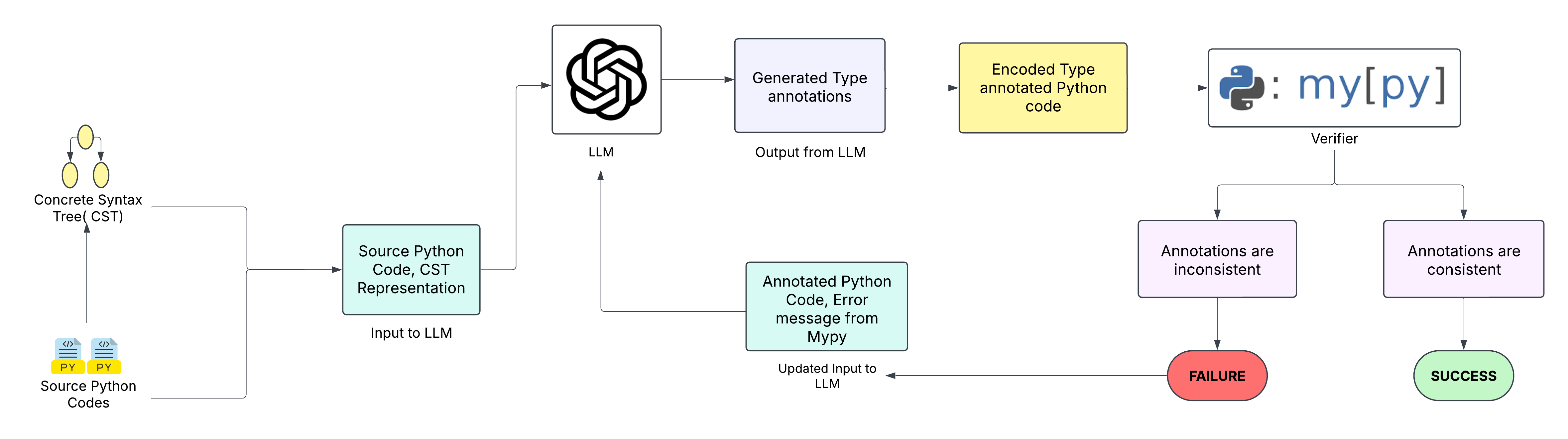}
  \caption{ Flow diagram of the generate–check–repair loop used for generation of type annotations}
  \label{fig:my_double_column_image}
\end{figure*}
\section{Introduction}
Dynamically typed languages such as Python have grown in popularity due to their flexibility, ease of use, and rapid prototyping capabilities. Python consistently ranks among the most popular programming languages in indices like IEEE Spectrum \cite{IEEE2024} and the TIOBE index \cite{TIOBE2024}. However, while dynamic typing facilitates quick development, it often leads to challenges in maintainability, error detection, and tooling support issues that static typing can help mitigate.

To address these challenges, the Python community introduced optional type annotations through PEP 484 \cite{PEP484}, which were later integrated into Python 3.5 along with static type checkers such as Mypy \cite{Mypy}. Empirical studies have shown that adding type annotations can prevent a significant portion of real‑world defects \cite{khan2021empirical} and that, once introduced, annotations tend to persist and uncover latent errors over time \cite{digrazia2022evolution}. However, manually annotating large codebases remains a manual‑intensive and error‑prone task.

Type annotation in Python began with PEP 484 \cite{PEP484}, leading to tools like Mypy \cite{Mypy}, PyType \cite{PyType}, PyRight \cite{PyRight}, and Pyre \cite{Pyre} for static type checking annotations in python. Early static analysis methods used rule-based inference and abstract interpretation but often defaulted to \texttt{Any} for complex cases. Learning-based approaches emerged to address these gaps: probabilistic models (e.g., JSNice \cite{JSNice}), graphical models \cite{XuPGM}, sequence models like DeepTyper \cite{DeepTyper} and NL2Type \cite{NL2Type}, and hybrids like TypeWriter \cite{Typewriter}. GNNs (e.g., LambdaNet \cite{Lambdanet}, OptTyper \cite{OptTyper}) and transformers (e.g., Typilus \cite{Typilus}, TypeBert \cite{TypeBert}) captured deeper semantic and structural patterns. Type4Py \cite{type4py} advanced this further by using deep similarity learning to cluster types in high-dimensional space, though all such methods remain constrained by training data quality.

The recent emergence of LLMs such as GPT‑3 \cite{brown2020language} and GPT‑4 \cite{gpt4}, trained on vast code and language corpora, offers open‑vocabulary, context‑aware type predictions. Reasoning‑optimized variants like O3-Mini \cite{O3Mini} and O4-Mini \cite{O4Mini}, with enhancements for logical inference and multi‑step problem solving, further strengthen this capability. However, to the best of our knowledge, existing researches and studies in this domain haven't comprehensively evaluated usage of LLMs to generate type annotations. This is the first work exploring the use of LLMs for generating type annotations.  

In this paper, we present a generate–check–repair pipeline for Python type annotation that tightly couples an LLM with the \texttt{Mypy} static checker. To give the model richer context about program structure and control flow, we first extract a Concrete Syntax Tree (CST) from the source code. The CST helps the LLM understand data and control dependencies, improving its ability to infer precise types. An initial prompt asks the LLM to produce a fully annotated version of the code; if \texttt{Mypy} reports no errors, the annotations are accepted. Otherwise, the exact error messages (with line numbers and expected vs.\ actual types) are fed back into the LLM in a repair prompt. This generate–check–repair cycle continues until \texttt{Mypy} passes or a fixed iteration limit is reached. We measure the effectiveness of our pipeline using two metrics : exact‑match accuracy (the percentage of annotations that exactly match the ground truth) and base‑type match accuracy (the percentage where the outer type constructor matches the reference, regardless of inner generics)~\cite{type4py}.

To validate our approach, we randomly sampled 6000 code snippets from the ManyTypes4Py benchmark~\cite{mt4py2021} and ran our generate–check–repair pipeline on this subset. We first measure the proportion of code snippets annotated by LLMs for which MyPy reported no errors (i.e., consistent results): GPT‑4o-Mini achieved consistency on 65.9\% of cases (34.1\% inconsistent), while GPT‑4.1‑mini, O3-Mini, and O4-Mini each reached approximately 88.6\% consistency (around 11.4\% failures). To measure annotation quality, we then compute exact‑match and base‑type match accuracies over all 6000 snippets in which GPT‑4o-Mini achieved 65.0\% exact‑match and 73.8\% base‑type accuracy, GPT‑4.1‑mini reached 70.5\% and 78.4\%, O3-Mini scored 70.2\% and 79.1\%, and O4-Mini obtained 68.2\% and 76.0\%. For comparison, Type4Py a deep‑learning model trained on the full ManyTypes4Py corpus reports 75.8\% exact‑match and 80.6\% base‑type accuracy. Although these figures come from a larger evaluation set, our LLMs with one-shot prompting close much of the gap without any task‑specific fine‑tuning or additional training, and crucially without requiring a labeled dataset. This demonstrates that general‑purpose and reasoning‑optimized LLMs can approximate state‑of‑the‑art type inference techniques, and that our pipeline can be adapted to other optionally typed languages.

\section{Related Work}
The evolution of type annotation in Python began with the introduction of optional type annotations in PEP 484 \cite{PEP484}. With the release of Python 3.5, these annotations initiated the addition of type annotations to existing Python codebases, leading to the development of various static type checkers, such as Mypy\cite{Mypy}. In addition to Mypy, several other tools have been developed, such as PyType \cite{PyType}, PyRight \cite{PyRight}, and Pyre \cite{Pyre} to provide early error detection and improve code maintainability by validating type annotations in dynamic languages.

Early work in Python type inference relied primarily on static analysis techniques. These methods use a predefined set of rules and constraints to infer types. Still, their precision is often limited by the dynamic nature of Python and the need to over approximate program behavior. Static-based approaches are effective for straightforward cases, but frequently struggle with complex language features and large codebases with numerous dependencies.

To overcome these limitations, researchers have explored learning based approaches for type inference. Early probabilistic models, such as JSNice \cite{JSNice}, employed conditional random fields to predict the identifier names and type annotations for JavaScript. Similarly, Xu et al. \cite{XuPGM} applied probabilistic graphical models to infer variable types in Python by capturing uncertain hints from attribute accesses, variable names, and data flows.

Deep learning techniques have further advanced the field. For example, DeepTyper \cite{DeepTyper} uses a sequence-to-sequence neural network to predict type annotations over an entire source file, while NL2Type \cite{NL2Type} leverages natural language information embedded in identifier names and comments to improve prediction accuracy. The TypeWriter model \cite{Typewriter} builds on these ideas by integrating deep neural networks with a combinatorial search strategy and external type checkers to validate its prediction of type.

Graph-based models have also emerged as a promising direction. LambdaNet \cite{Lambdanet} constructs type dependency graphs to integrate logical constraints and contextual clues, and OptTyper \cite{OptTyper} formulates type inference as an optimization problem that combines deterministic type system rules with natural language features. More recently, Typilus \cite{Typilus} and TypeBert \cite{TypeBert} have pushed the state of the art by employing graph neural networks and transformer models, respectively, to overcome the limitations imposed by a fixed-type vocabulary. A notable advancement in learning-based type inference is Type4Py \cite{type4py}. In contrast to models constrained by a fixed-type vocabulary, Type4Py leverages a deep similarity learning framework that projects code elements into a high-dimensional feature space, effectively clustering similar types.

\textit{Building on these efforts, our work departs from both rule‑driven static analysis and data‑intensive neural models by employing open-vocabulary LLMs within a generate–check–repair loop driven by a static type checker. This design avoids fixed type vocabularies and large labeled training corpora while still guaranteeing soundness through verification. By evaluating general‑purpose and reasoning optimized LLMs on a shared benchmark, we show that such models can approach the accuracy of specialized tools like Type4Py, current state of the art deep-learning based tool and that the approach is readily transferable to other gradually typed languages.}

\begin{figure*}[t]
  \centering
  \includegraphics[width=\textwidth]{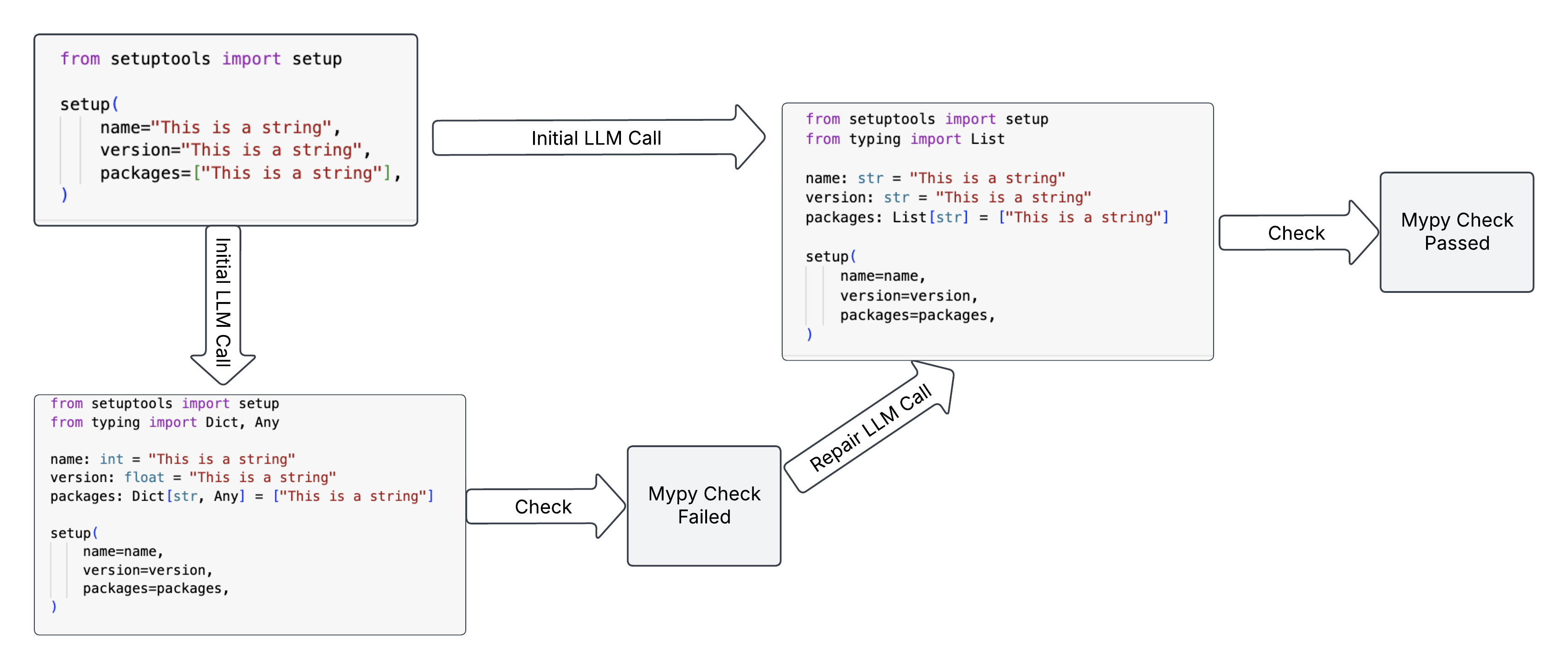}
  \caption{High‑level walkthrough of our generate–check–repair pipeline on an example code snippet. The top row illustrates a case where the LLM’s initial annotation  is sent to \texttt{Mypy} and passes on the first try. The bottom row shows the case where the LLM’s initial annotation  is sent to \texttt{Mypy} fails, and is then refined via a repair prompt to produce a correct annotation . Note that CST extraction and prompt formatting occur before these steps; this diagram focuses on the core annotation, verification, and repair loop.}

  \label{fig:my_double_column_image1}
\end{figure*}

\section{Proposed Methodology}

In this section, we describe our automated pipeline for Python type annotation, which combines three core components: (1) data preprocessing to extract and prepare context, (2) LLM-based annotation generation and repair, and (3) static verification via \texttt{Mypy}. Figure~\ref{fig:my_double_column_image} illustrates the end‑to‑end flow of this pipeline.

\subsection{Data Preprocessing}
We begin by parsing each unannotated Python file into a Concrete Syntax Tree (CST) using \texttt{libcst} Python module. The CST preserves the full syntactic structure of the code including comments, doc‑strings, and formatting, ensuring that no semantic or contextual information is lost. This richer context capturing control‑flow and data‑flow relationships enables the model to make more informed type predictions than using raw text alone.

Although our prototype uses a Python-specific \texttt{libcst} frontend to generate the CST, the approach is language agnostic i.e any source language for which a CST can be produced can plug into the same pipeline. Various parsers exist for other languages. For example, the Chevrotain Java parser emits CSTs for Java~\cite{chevrotainjava}, ANTLR grammars generate concrete (parse) trees for C and many other languages~\cite{antlrC,antlrparsetree}, and Tree‑sitter builds lossless CSTs across a wide range of languages~\cite{treesitter}. In practice, only the CST extraction component needs to be swapped to target a different language; the downstream LLM prompting and \texttt{Mypy}-style checking loop remain unchanged.

Once the CST is extracted, we assemble the initial prompt for the LLM. The prompt includes (a) the original Python code, (b) CST representation of the code, and (c) a clear instruction to insert type annotations for all functions without altering any other code. We carefully format this prompt to minimize noise and guide the model toward outputting only the annotated code. The exact prompt templates we use for both the initial annotation and any subsequent repair calls are detailed in Section \ref{sec:annotation_generation} below.

\subsection{Annotation Generation and Repair}
\label{sec:annotation_generation}
Our LLM component is invoked twice per iteration: once to generate initial type annotations and again to repair it when needed.

\paragraph{\textbf{Initial Annotation}} : 
An initial prompt presents the unannotated code and CST, requesting the model to produce a type‑annotated version of the code. The prompt used is as follows : 

\medskip
\textit{You are a Python code analysis assistant. Your task is to analyze a given Python code snippet along with its corresponding Concrete Syntax Tree (CST). Using this information:}

\begin{itemize}
    \item \textit{Identify the types of all variables, function arguments, and return values based on the code and its structure.}
    \item \textit{Generate an updated type-annotated version of the Python code.}
    \item \textit{Ensure the type annotations are precise and conform to Python's type hinting standards.}
    \item \textit{Include any inferred types for variables or functions not explicitly annotated in the original code.}
\end{itemize}
\textit{Input:}
\begin{itemize}
    \item \textit{Python Code: The source code of a Python program.}
    \item \textit{CST Representation: The Concrete Syntax Tree corresponding to the code.}
\end{itemize}
\textit{Output:}
\begin{itemize}
    \item \textit{The type-annotated version of the input Python code.}
\end{itemize}
\textit{Note: Ensure the output is syntactically correct and can be validated by Mypy. If there are ambiguous types, make reasonable assumptions and include them in the annotations.The type must be inferred in the context of the entire Python program}

\textit{For example, for the given input code : \{ EXAMPLE CODE \}  and its corresponding CST representation: \{ EXAMPLE CST REPRESENTATION \}}

\textit{The output should be as follows : \{ EXAMPLE ANNOTATED PYTHON CODE \}}

\textit{I just want you to return me the updated code which should be type annotated. There is no need to return any reasoning for the same or any other comments. Just return the updated code.}

\medskip
The LLM generates candidate annotations for function signatures, variable declarations, and return types.

\paragraph{\textbf{Repair Annotation}} :
If the candidate fails verification from Mypy, we extract the exact \texttt{Mypy} errors listing line numbers, expected types, and actual types and feed them back into a second “repair” prompt. The prompt used is as follows :  

\medskip
\textit{You are a Python code analysis assistant. Your task is to analyze a given annotated Python code snippet along with its corresponding errors which has been generated by checking the given Python code using Mypy. Using this information:}

\begin{itemize}
    \item \textit{Identify the types of all variables, function arguments, and return values based on the code and its structure.}
    \item \textit{Generate an updated type-annotated version of the Python code.}
    \item \textit{Ensure the type annotations are precise and conform to Python's type hinting standards.}
    \item \textit{Include any inferred types for variables or functions not explicitly annotated in the original code.}
\end{itemize}

\textit{Input:}
\begin{itemize}
    \item \textit{Python Code: The annotated code of a Python program.}
    \item \textit{Error message: The corresponding errors generated while checking the Python code using Mypy .}
\end{itemize}

Output:
\begin{itemize}
    \item \textit{The updated type-annotated version of the input Python code.}
\end{itemize}
\textit{Note: Ensure the output is syntactically correct and can be validated by Mypy. If there are ambiguous types, make reasonable assumptions and include them in the annotations.The type must be inferred in the context of the entire Python program}

\textit{For example, for the given input code : \{ EXAMPLE CODE \} and its corresponding error message: \{ EXAMPLE ERROR MESSAGE \}}

\textit{After analysing the error, The output should be as follows :
\{ CORRECTED OUTPUT CODE \}}

\textit{I just want you to return me the updated code which should be type annotated. There is no need to return any reasoning for the same or any other comments. Just return the updated code.}

\medskip
This prompt re‑provides the last annotated code and asks the model to correct the specific mismatches. The LLM refines its annotations based on this targeted feedback. These generate–check–repair calls repeat until either \texttt{Mypy} passes or a predefined iteration limit is reached.

\subsection{Verification with \texttt{Mypy}}

Each candidate annotation produced by the LLM is checked using a Python helper routine that invokes \texttt{Mypy} as an external process. Specifically, we write the annotated code to a temporary file and run:

\begin{lstlisting}[language=bash]
mypy --install-types --non-interactive 
      --ignore-missing-imports 
      --follow-imports=silent temp_code.py
\end{lstlisting}

The flags ensure that any required stub packages are installed automatically (\texttt{install-types}), the call does not prompt for user input (\texttt{non-interactive}), and missing imports do not halt the check (\texttt{ignore-missing-imports}, \texttt{follow-imports=silent}).These flags prevent unresolved imports from causing false positives when code snippets reference classes or functions defined elsewhere in a larger codebase that isn’t available during checking. We capture both \texttt{stdout} and \texttt{stderr}, and interpret a zero exit code as “no type errors.” Otherwise, we aggregate the full error output into an in‑memory string and return a failure signal.

If \texttt{Mypy} passes, the pipeline accepts the annotations and terminates. If it fails, the collected error messages including line numbers, expected vs.\ actual types, and any import issues are supplied back to the LLM in the repair prompt. Figure~\ref{fig:my_double_column_image1} illustrates the running example of this pipeline 
\section{Experimental Setup and Results}

\subsection{Dataset}

 For our experiments, we used the ManyTypes4Py dataset~\cite{mt4py2021} which consists of partially annotated Python programs. To curate the dataset, the authors of ManyTypes4Py selected open source Python projects that explicitly list Mypy as a dependency, as its usage is a strong indicator that the code contains type annotations. This filtering process resulted in 5,382 high-quality projects from GitHub which were then de-duplicated using a tool to reduce duplication bias. The correctness of the type annotations in these projects was then verified by Mypy. Each of these project contains various program snippets. We conducted our experiments on a subset of 6000 randomly sampled Python programs snippets to evaluate the effectiveness of our pipeline.  

\subsection{Models}

We tested our pipeline against two different categories of LLMs.

\paragraph{General‑Purpose LLMs}
\begin{itemize}
  \item GPT‑4o Mini~\cite{openai2025gpt4omini} is a compact variant of GPT‑4, optimized for efficient code comprehension and generation.
  \item GPT‑4.1‑mini~\cite{openai2024gpt4_1mini} is a scaled‑down GPT‑4.1 model that balances performance with lower inference latency.
\end{itemize}

\paragraph{Reasoning‑Optimized LLMs}
\begin{itemize}
  \item O3-Mini~\cite{O3Mini} is trained on reasoning‑heavy data to excel at multi‑step logical inferences.
  \item O4-Mini~\cite{O4Mini} extends O3-Mini with enhanced context handling and improved consistency on complex tasks.
\end{itemize}

We set the sampling temperature to 0.7 for all LLM queries and imposed no maximum token limit, allowing each model to leverage its full context window. This allowed them to annotate large code blocks as well. 

Based on observations from prior work involving LLM based loops, it was observed that loops can stagnate after many repair iterations~\cite{wu2024lemur, kamath2023finding}, we experimented with various iteration limits and chose \(N=10\) as a balance between refinement capacity and avoiding redundant proposals.

To assess annotation quality against the ground‑truth labels provided by ManyTypes4Py, we first convert each LLM’s output into the dataset’s JSON schema using Libsa4Py~\cite{libsa4py,manytypes4py_github}. This enables direct comparison with the reference annotations and the computation of exact‑match and base‑type metrics as defined by the Type4Py benchmark. Due to minor syntax issues, about 150 snippets could not be converted and were therefore excluded from these calculations.

\subsection{Evaluation Metrics}
\label{sec:metrics}
We our pipeline using two key metrics:

\begin{itemize}
  \item[EM1.] \textbf{Static Consistency.} What fraction of generated annotations pass \texttt{Mypy} with zero errors i.e consistent results. We measure this consistency as the percentage of snippets for which the pipeline produces code annotations such that no type errors are reported.

  \item[EM2.] \textbf{Annotation Quality.} How well do the generated type annotations align with the ground‑truth labels which are given in ManyTypes4Py dataset. We quantify this quality using two established metrics from the Type4Py evaluation~\cite{type4py}:
    \begin{itemize}
      \item \textbf{Exact Match Accuracy:} The percentage of annotations where the model’s output exactly equals the ground truth type annotation as present in the dataset.
      \item \textbf{Base Type Match Accuracy:} The percentage of annotations where the outermost type constructor matches the ground truth, ignoring differences in its parameters (for example, both \texttt{List[str]} and \texttt{List[int]} count as a correct base‑type match because they share the \texttt{List} constructor).
    \end{itemize}
\end{itemize}

Both metrics are computed independently over the full set of 6000 snippets, regardless of whether each snippet passed the Mypy consistency check.

\subsection{Results}  
We organize our evaluation along the two metrics detailed in Section \ref{sec:metrics} :
(1) \emph{Static Consistency}, the fraction of snippets whose annotations pass \texttt{mypy} with zero errors, and  
(2) \emph{Annotation Quality}, measured by exact‑match and base‑type accuracies against the ManyTypes4Py ground truth. 

Using the static consistency metric, we find that GPT‑4o-Mini produced annotations that passed \texttt{Mypy} without errors on 65.9\% of the 6000 snippets while failing on 34.1\% i.e inconsistent results, whereas GPT‑4.1‑mini, O3 Mini, and O4 Mini each achieve approximately 88.6\% consistency while failing for 11.4\% of code snippets.

Next, we assess annotation quality over the full 6000‑snippet set using the exact‑match and base‑type metrics defined earlier. Table~\ref{table:accuracy_comparison} summarizes these results:

\begin{table}[h]
    \centering
    \caption{Exact and Base‑Type Match Accuracy (\%)}
    \begin{tabular}{|l|c|c|}
        \hline
        \textbf{Model} & \textbf{Exact Match} & \textbf{Base‑Type Match} \\
        \hline
        GPT‑4o Mini         & 65.0 & 73.8 \\
        GPT‑4.1 Mini        & 70.5 & 78.4 \\
        O3 Mini             & 70.2 & 79.1 \\
        O4 Mini             & 68.2 & 76.0 \\
        \hline
    \end{tabular}
    \label{table:accuracy_comparison}
\end{table}

\paragraph{\textbf{Quantitative Analysis}}: The four LLMs differ in their accuracy to generate the type annotations. GPT-4o Mini achieves 67.2\% exact-match and 75.8\% base-type accuracy, while GPT-4.1-mini improves to 70.5\% and 78.4 \%. Among the reasoning-optimized models, O3 Mini reaches 70.2\% exact-match and 79.1\% base-type accuracy, and O4 Mini attains 68.2\% and 76.0\%. 

To get a deeper understanding of the performance, we track two efficiency metrics: (1) the average number of repair iterations required for snippets that ultimately pass \texttt{Mypy} within our iteration limit, and (2) the proportion of those successful snippets resolved on the first attempt. GPT‑4o-Mini requires 0.78 repair rounds on average and succeeds without any repair in 86.4\% of converged cases. By comparison, GPT‑4.1‑mini, O3 Mini, and O4 Mini each average roughly 0.54 repair iterations and resolve over 91\% of code snippets on their initial proposal with O4-Mini slightly outperforming the others. This demonstrates that both the latest general‑purpose and the reasoning‑optimized LLMs achieve rapid, high‑quality annotations when guided by precise \texttt{Mypy} feedback.

Type4Py in general, generates a ranked list of possible types for a particular variable which isn't the case for LLMs. LLMs only give one type for each variable. Hence, we only consider the top-1 prediction for a fair comparison with LLMs. On the complete ManyTypes4Py benchmark, Type4Py achieves 75.8\% exact‑match and 80.6\% base‑type accuracy. Although these figures are measured over a larger corpus, our LLMs with one-shot prompting evaluated on the 6000 code snippet subset from the same benchmark on which Type4py has been evaluated nearly close the gap, demonstrating that they deliver competitive annotation quality without any task‑specific training.

\paragraph{\textbf{Qualitative Analysis}} : After examining failure cases and successful annotations, we identify the following key qualitative insights:

\smallskip  
\noindent\textbf{Excessively large code snippets} 

When processing large code snippets longer than 150 lines along with their CST representation, the model struggles to generate type annotations. In these situations, earlier definitions and control‑flow relationships may have become inaccessible mid‑generation, leading the model to produce annotations that were plausible in isolation but inconsistent when viewed across the entire snippet.

To mitigate these failures, we can first split large files into smaller, self‑contained units involving individual functions or classes along with only their direct dependencies. Each unit could be annotated in isolation, with its signature and a minimal CST fragment.

\smallskip  
\noindent\textbf{Failure cases beyond incorrect annotation} 

A substantial fraction of “inconsistent” cases were not due to incorrect annotations but because of unrelated code issues. Common \texttt{Mypy} complaints included arguments not being converted properly during string formatting, missing return statements in the code, invalid syntax, undefined names and failure in performing relative imports. For example, \texttt{Mypy} would report:

\begin{lstlisting}[basicstyle=\ttfamily\small]
The code has annotation errors:
Found errors in 1 file (checked 1 source file)

temp_code.py:12: error: Not all arguments 
converted during string formatting  [str-format]

temp_code.py:143: error: Missing return statement  [return]

temp_code.py:123: error: Name "t" is not defined 
[name-defined]

temp_code.py:24: error: No parent module  
cannot perform relative import

temp_code.py:372: error: Name "root" already
defined on line 370  [no-redef]
\end{lstlisting}

To mitigate these failures, running a quick syntax‑only lint such as flake8 \cite{flake8} can catch and correct formatting, missing returns, and undefined names before annotation. Stripping or ignoring non‑type Mypy errors such as string‑format or redefinition warnings, 
simple import stubs or path adjustments can resolve relative‑import errors thereby allowing the pipeline to focus purely on inferring and refining type annotations.

\smallskip  
\noindent\textbf{Generic annotations} 

When presented with unfamiliar code snippets, the LLM sometimes defaulted to overly broad annotations. This can be seen in the following example : 

\begin{figure}[ht]
  \centering
  \includegraphics[width=\columnwidth]{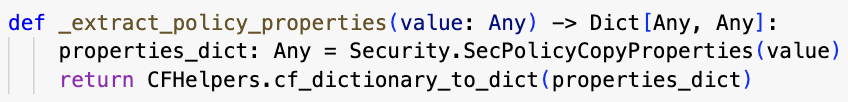}
  \caption{Example of overly broad LLM annotation where the LLM has generated \texttt{Dict[Any, Any]} for the return type and also defined Any for the variable inside the function}
  \label{fig:llm_annotation_examples}
\end{figure}

To mitigate these failures, we can scan for generic placeholders such as \texttt{Any} and automatically flag them for a targeted repair prompt.

\section{Conclusion and Future Work}

Dynamically typed languages like Python enable rapid development but often at the expense of tooling support and early bug detection. Adding type annotations to code helps in increasing the robustness.

In this work, we tackled the challenge of automating type annotations by introducing a generate–check–repair pipeline that combines large language models with the \texttt{Mypy} static checker. By enriching each prompt with a Concrete Syntax Tree to capture control and data flow, and by feeding precise \texttt{Mypy} error messages back into the model for correction, our approach ensures that all accepted annotations are fully consistent with the code. 

We evaluated four LLM variants on a 6000 python code snippets subset of the ManyTypes4Py benchmark. We first measure the proportion of code snippets annotated by LLMs for which MyPy reported no errors (i.e., consistent results): GPT‑4o-Mini achieved consistency on 65.9\% of cases (34.1\% inconsistent), while GPT‑4.1‑mini, O3-Mini, and O4-Mini each reached approximately 88.6\% consistency (around 11.4\% failures). To measure annotation quality, we then compute exact‑match and base‑type match accuracies over all 6000 snippets in which GPT‑4.1‑mini and O3 Mini led the pack, reaching 70.5\% exact‑match and 79.1\% base‑type accuracy in a one‑shot setting. For comparison, Type4Py reports 75.8\% and 80.6\% on the full corpus measurements that cover a larger sample but still highlight how closely our models follow a specialized, trained system. Moreover, both GPT‑4.1‑mini and O3-Mini required under one repair iteration on average demonstrating rapid and robust annotation under our generate–check–repair pipeline.  These findings show that even in a one‑shot setting and without any task‑specific fine‑tuning, both general‑purpose and reasoning‑optimized LLMs can generate accurate, high‑quality type annotations. LLMs can be readily adapted to other languages with optional or gradual typing to generate type annotations without the need to develop specific tools or amass large labeled datasets for model training.

 We plan to extend this paradigm in several directions. Firstly, we aim to fine‑tune both general‑purpose and reasoning‑optimized LLMs on language‑specific codebases to enhance their ability to infer complex and user‑defined types which may boost annotation accuracy in specific domains. We also plan to integrate our generate–check pipeline with existing type‑inference tools, combining statistical and rule‑based analyses with LLM adaptability to create a hybrid framework that is both robust and reliable for comprehensive type annotation. Finally, we believe this approach can generalize to other languages with optional or gradual typing (e.g., JavaScript, Ruby) and integrate seamlessly into development tools, providing on‑the‑fly type suggestions and corrections. By marrying the generative prowess of LLMs with the rigor of static verification, we lay the groundwork for more reliable, maintainable code in dynamic language ecosystems.


\bibliographystyle{ACM-Reference-Format}
\bibliography{base}


\begin{thebibliography}{35}

%%% the Web address.

\ifx \showCODEN    \undefined \def \showCODEN     #1{\unskip}     \fi
\ifx \showDOI      \undefined \def \showDOI       #1{#1}\fi
\ifx \showISBNx    \undefined \def \showISBNx     #1{\unskip}     \fi
\ifx \showISBNxiii \undefined \def \showISBNxiii  #1{\unskip}     \fi
\ifx \showISSN     \undefined \def \showISSN      #1{\unskip}     \fi
\ifx \showLCCN     \undefined \def \showLCCN      #1{\unskip}     \fi
\ifx \shownote     \undefined \def \shownote      #1{#1}          \fi
\ifx \showarticletitle \undefined \def \showarticletitle #1{#1}   \fi
\ifx \showURL      \undefined \def \showURL       {\relax}        \fi
% invisible to TeX
\providecommand\bibfield[2]{#2}
\providecommand\bibinfo[2]{#2}
\providecommand\natexlab[1]{#1}
\providecommand\showeprint[2][]{arXiv:#2}

\bibitem[Achiam et~al\mbox{.}(2023)]%
        {gpt4}
\bibfield{author}{\bibinfo{person}{Josh Achiam}, \bibinfo{person}{Steven Adler}, \bibinfo{person}{Sandhini Agarwal}, \bibinfo{person}{Lama Ahmad}, \bibinfo{person}{Ilge Akkaya}, \bibinfo{person}{Florencia~Leoni Aleman}, \bibinfo{person}{Diogo Almeida}, \bibinfo{person}{Janko Altenschmidt}, \bibinfo{person}{Sam Altman}, \bibinfo{person}{Shyamal Anadkat}, {et~al\mbox{.}}} \bibinfo{year}{2023}\natexlab{}.
\newblock \showarticletitle{Gpt-4 technical report}.
\newblock \bibinfo{journal}{\emph{arXiv preprint arXiv:2303.08774}} (\bibinfo{year}{2023}).
\newblock
\urldef\tempurl%
\url{https://arxiv.org/abs/2303.08774}
\showURL{%
\tempurl}


\bibitem[Allamanis et~al\mbox{.}(2020)]%
        {Typilus}
\bibfield{author}{\bibinfo{person}{Miltiadis Allamanis}, \bibinfo{person}{Earl~T. Barr}, \bibinfo{person}{Soline Ducousso}, {and} \bibinfo{person}{Zheng Gao}.} \bibinfo{year}{2020}\natexlab{}.
\newblock \showarticletitle{Typilus: neural type hints}. In \bibinfo{booktitle}{\emph{Proceedings of the 41st ACM SIGPLAN Conference on Programming Language Design and Implementation}} \emph{(\bibinfo{series}{PLDI ’20})}. \bibinfo{publisher}{ACM}, \bibinfo{pages}{91–105}.
\newblock
\urldef\tempurl%
\url{https://doi.org/10.1145/3385412.3385997}
\showDOI{\tempurl}


\bibitem[Brown et~al\mbox{.}(2020)]%
        {brown2020language}
\bibfield{author}{\bibinfo{person}{Tom Brown}, \bibinfo{person}{Benjamin Mann}, \bibinfo{person}{Nick Ryder}, \bibinfo{person}{Melanie Subbiah}, \bibinfo{person}{Jared~D Kaplan}, \bibinfo{person}{Prafulla Dhariwal}, \bibinfo{person}{Arvind Neelakantan}, \bibinfo{person}{Pranav Shyam}, \bibinfo{person}{Girish Sastry}, \bibinfo{person}{Amanda Askell}, {et~al\mbox{.}}} \bibinfo{year}{2020}\natexlab{}.
\newblock \showarticletitle{Language models are few-shot learners}.
\newblock \bibinfo{journal}{\emph{Advances in neural information processing systems}}  \bibinfo{volume}{33} (\bibinfo{year}{2020}), \bibinfo{pages}{1877--1901}.
\newblock
\urldef\tempurl%
\url{https://arxiv.org/abs/2005.14165}
\showURL{%
\tempurl}


\bibitem[Community(2016)]%
        {antlrC}
\bibfield{author}{\bibinfo{person}{ANTLR Community}.} \bibinfo{year}{2016}\natexlab{}.
\newblock \bibinfo{title}{C Grammars for ANTLR v4 (ANSI C / GNU C)}.
\newblock \bibinfo{howpublished}{\url{https://github.com/antlr/grammars-v4/tree/master/c}}.
\newblock
\newblock
\shownote{ANTLR grammars for C that yield parse/CST trees}.


\bibitem[Developers(2012)]%
        {Mypy}
\bibfield{author}{\bibinfo{person}{Mypy Developers}.} \bibinfo{year}{2012}\natexlab{}.
\newblock \bibinfo{title}{Mypy: Optional Static Type Checking for Python}.
\newblock \bibinfo{howpublished}{\url{https://mypy-lang.org/}}.
\newblock


\bibitem[Di~Grazi{\'a} and Pradel(2022)]%
        {digrazia2022evolution}
\bibfield{author}{\bibinfo{person}{Luca Di~Grazi{\'a}} {and} \bibinfo{person}{Michael Pradel}.} \bibinfo{year}{2022}\natexlab{}.
\newblock \showarticletitle{The Evolution of Type Annotations in Python: An Empirical Study}. In \bibinfo{booktitle}{\emph{Proceedings of the 30th ACM Joint European Software Engineering Conference and Symposium on the Foundations of Software Engineering}} \emph{(\bibinfo{series}{ESEC/FSE ’22})}. \bibinfo{pages}{209--220}.
\newblock
\urldef\tempurl%
\url{https://doi.org/10.1145/3540250.3549114}
\showDOI{\tempurl}


\bibitem[Facebook(nd)]%
        {Pyre}
\bibfield{author}{\bibinfo{person}{Facebook}.} \bibinfo{year}{n.d.}\natexlab{}.
\newblock \bibinfo{title}{Pyre: A Performant Type Checker for Python}.
\newblock \bibinfo{howpublished}{\url{https://pyre-check.org}}.
\newblock


\bibitem[Google(nd)]%
        {PyType}
\bibfield{author}{\bibinfo{person}{Google}.} \bibinfo{year}{n.d.}\natexlab{}.
\newblock \bibinfo{title}{PyType: A Python Type Checker}.
\newblock \bibinfo{howpublished}{\url{https://github.com/google/pytype}}.
\newblock


\bibitem[Hellendoorn et~al\mbox{.}(2018)]%
        {DeepTyper}
\bibfield{author}{\bibinfo{person}{Vincent~J Hellendoorn}, \bibinfo{person}{Christian Bird}, \bibinfo{person}{Earl~T Barr}, {and} \bibinfo{person}{Miltiadis Allamanis}.} \bibinfo{year}{2018}\natexlab{}.
\newblock \showarticletitle{Deep learning type inference}. In \bibinfo{booktitle}{\emph{Proceedings of the 2018 26th acm joint meeting on european software engineering conference and symposium on the foundations of software engineering}}. \bibinfo{pages}{152--162}.
\newblock
\urldef\tempurl%
\url{https://doi.org/10.1145/3236024.3236051}
\showURL{%
\tempurl}


\bibitem[{IEEE Spectrum}(2024)]%
        {IEEE2024}
\bibfield{author}{\bibinfo{person}{{IEEE Spectrum}}.} \bibinfo{year}{2024}\natexlab{}.
\newblock \bibinfo{title}{Top Programming Languages 2024}.
\newblock \bibinfo{howpublished}{\url{http://spectrum.ieee.org/top-programming-languages-2024}}.
\newblock


\bibitem[Jesse et~al\mbox{.}(2021)]%
        {TypeBert}
\bibfield{author}{\bibinfo{person}{Kevin Jesse}, \bibinfo{person}{Premkumar~T. Devanbu}, {and} \bibinfo{person}{Toufique Ahmed}.} \bibinfo{year}{2021}\natexlab{}.
\newblock \showarticletitle{Learning type annotation: is big data enough?} \emph{(\bibinfo{series}{ESEC/FSE 2021})}. \bibinfo{publisher}{Association for Computing Machinery}, \bibinfo{address}{New York, NY, USA}, \bibinfo{pages}{1483–1486}.
\newblock
\showISBNx{9781450385626}
\urldef\tempurl%
\url{https://doi.org/10.1145/3468264.3473135}
\showDOI{\tempurl}


\bibitem[Kamath et~al\mbox{.}(2023)]%
        {kamath2023finding}
\bibfield{author}{\bibinfo{person}{A. Kamath}, \bibinfo{person}{A. Senthilnathan}, \bibinfo{person}{S. Chakraborty}, \bibinfo{person}{P. Deligiannis}, \bibinfo{person}{S. Lahiri}, \bibinfo{person}{A. Lal}, \bibinfo{person}{A. Rastogi}, \bibinfo{person}{S. Roy}, {and} \bibinfo{person}{R. Sharma}.} \bibinfo{year}{2023}\natexlab{}.
\newblock \showarticletitle{Finding inductive loop invariants using large language models}. In \bibinfo{booktitle}{\emph{arXiv preprint}}.
\newblock
\urldef\tempurl%
\url{https://arxiv.org/abs/2311.07948}
\showURL{%
\tempurl}


\bibitem[Khan et~al\mbox{.}(2022)]%
        {khan2021empirical}
\bibfield{author}{\bibinfo{person}{Faizan Khan}, \bibinfo{person}{Boqi Chen}, \bibinfo{person}{Daniel Varro}, {and} \bibinfo{person}{Shane McIntosh}.} \bibinfo{year}{2022}\natexlab{}.
\newblock \showarticletitle{An Empirical Study of Type-Related Defects in Python Projects}.
\newblock \bibinfo{journal}{\emph{IEEE Transactions on Software Engineering}} \bibinfo{volume}{48}, \bibinfo{number}{8} (\bibinfo{year}{2022}), \bibinfo{pages}{3145--3158}.
\newblock
\urldef\tempurl%
\url{https://doi.org/10.1109/TSE.2021.3082068}
\showDOI{\tempurl}


\bibitem[Malik et~al\mbox{.}(2019)]%
        {NL2Type}
\bibfield{author}{\bibinfo{person}{Rabee~Sohail Malik}, \bibinfo{person}{Jibesh Patra}, {and} \bibinfo{person}{Michael Pradel}.} \bibinfo{year}{2019}\natexlab{}.
\newblock \showarticletitle{NL2Type: Inferring JavaScript Function Types from Natural Language Information}. In \bibinfo{booktitle}{\emph{2019 IEEE/ACM 41st International Conference on Software Engineering (ICSE)}}. \bibinfo{pages}{304--315}.
\newblock
\urldef\tempurl%
\url{https://doi.org/10.1109/ICSE.2019.00045}
\showDOI{\tempurl}


\bibitem[Microsoft(nd)]%
        {PyRight}
\bibfield{author}{\bibinfo{person}{Microsoft}.} \bibinfo{year}{n.d.}\natexlab{}.
\newblock \bibinfo{title}{PyRight: A Static Type Checker for Python}.
\newblock \bibinfo{howpublished}{\url{https://github.com/microsoft/pyright}}.
\newblock


\bibitem[Mir et~al\mbox{.}(2021)]%
        {mt4py2021}
\bibfield{author}{\bibinfo{person}{A.~M. Mir}, \bibinfo{person}{E. Latoskinas}, {and} \bibinfo{person}{G. Gousios}.} \bibinfo{year}{2021}\natexlab{}.
\newblock \showarticletitle{ManyTypes4Py: A Benchmark Python Dataset for Machine Learning-Based Type Inference}. In \bibinfo{booktitle}{\emph{IEEE/ACM 18th International Conference on Mining Software Repositories (MSR)}}. \bibinfo{publisher}{IEEE Computer Society}, \bibinfo{pages}{585--589}.
\newblock
\urldef\tempurl%
\url{https://doi.org/10.1109/MSR52588.2021.00079}
\showDOI{\tempurl}


\bibitem[Mir et~al\mbox{.}(2022)]%
        {type4py}
\bibfield{author}{\bibinfo{person}{Amir~M. Mir}, \bibinfo{person}{Evaldas Latoškinas}, \bibinfo{person}{Sebastian Proksch}, {and} \bibinfo{person}{Georgios Gousios}.} \bibinfo{year}{2022}\natexlab{}.
\newblock \showarticletitle{Type4Py: practical deep similarity learning-based type inference for python}. In \bibinfo{booktitle}{\emph{Proceedings of the 44th International Conference on Software Engineering}} \emph{(\bibinfo{series}{ICSE ’22})}. \bibinfo{publisher}{ACM}.
\newblock
\urldef\tempurl%
\url{https://doi.org/10.1145/3510003.3510124}
\showDOI{\tempurl}


\bibitem[OpenAI(2024)]%
        {openai2024gpt4_1mini}
\bibfield{author}{\bibinfo{person}{OpenAI}.} \bibinfo{year}{2024}\natexlab{}.
\newblock \bibinfo{title}{{GPT-4.1-mini Model}}.
\newblock \bibinfo{howpublished}{\url{https://openai.com/index/gpt-4-1/}}.
\newblock


\bibitem[OpenAI(2025a)]%
        {openai2025gpt4omini}
\bibfield{author}{\bibinfo{person}{OpenAI}.} \bibinfo{year}{2025}\natexlab{a}.
\newblock \bibinfo{title}{{GPT-4o Mini Model}}.
\newblock \bibinfo{howpublished}{\url{https://platform.openai.com/docs/models/gpt-4o-mini}}.
\newblock


\bibitem[OpenAI(2025b)]%
        {O3Mini}
\bibfield{author}{\bibinfo{person}{OpenAI}.} \bibinfo{year}{2025}\natexlab{b}.
\newblock \bibinfo{title}{GPT O3 mini Model}.
\newblock \bibinfo{howpublished}{\url{https://openai.com/index/openai-o3-mini/}}.
\newblock


\bibitem[OpenAI(2025c)]%
        {O4Mini}
\bibfield{author}{\bibinfo{person}{OpenAI}.} \bibinfo{year}{2025}\natexlab{c}.
\newblock \bibinfo{title}{GPT o4 mini Model}.
\newblock \bibinfo{howpublished}{\url{https://openai.com/index/introducing-o3-and-o4-mini/}}.
\newblock


\bibitem[Pandi et~al\mbox{.}(2021)]%
        {OptTyper}
\bibfield{author}{\bibinfo{person}{Irene~Vlassi Pandi}, \bibinfo{person}{Earl~T. Barr}, \bibinfo{person}{Andrew~D. Gordon}, {and} \bibinfo{person}{Charles Sutton}.} \bibinfo{year}{2021}\natexlab{}.
\newblock \bibinfo{title}{OptTyper: Probabilistic Type Inference by Optimising Logical and Natural Constraints}.
\newblock
\newblock
\showeprint[arxiv]{2004.00348}~[cs.PL]
\urldef\tempurl%
\url{https://arxiv.org/abs/2004.00348}
\showURL{%
\tempurl}


\bibitem[Parr(2013)]%
        {antlrparsetree}
\bibfield{author}{\bibinfo{person}{Terence Parr}.} \bibinfo{year}{2013}\natexlab{}.
\newblock \bibinfo{title}{ANTLR v4 Parse Trees (Concrete Syntax Trees)}.
\newblock \bibinfo{howpublished}{\url{https://github.com/antlr/antlr4/blob/master/doc/faq/general.md}}.
\newblock
\newblock
\shownote{ANTLR produces full parse trees that correspond to CSTs}.


\bibitem[Pradel et~al\mbox{.}(2020)]%
        {Typewriter}
\bibfield{author}{\bibinfo{person}{Michael Pradel}, \bibinfo{person}{Georgios Gousios}, \bibinfo{person}{Jason Liu}, {and} \bibinfo{person}{Satish Chandra}.} \bibinfo{year}{2020}\natexlab{}.
\newblock \bibinfo{title}{TypeWriter: Neural Type Prediction with Search-based Validation}.
\newblock
\newblock
\showeprint[arxiv]{1912.03768}~[cs.SE]
\urldef\tempurl%
\url{https://arxiv.org/abs/1912.03768}
\showURL{%
\tempurl}


\bibitem[{PyCQA}(2024)]%
        {flake8}
\bibfield{author}{\bibinfo{person}{{PyCQA}}.} \bibinfo{year}{2024}\natexlab{}.
\newblock \bibinfo{title}{{flake8}: the modular source code checker for Python}.
\newblock
\newblock
\urldef\tempurl%
\url{https://flake8.pycqa.org/}
\showURL{%
\tempurl}


\bibitem[{Python Software Foundation}(2014)]%
        {PEP484}
\bibfield{author}{\bibinfo{person}{{Python Software Foundation}}.} \bibinfo{year}{2014}\natexlab{}.
\newblock \bibinfo{title}{PEP 484 -- Type Hints}.
\newblock \bibinfo{howpublished}{\url{https://peps.python.org/pep-0484/}}.
\newblock


\bibitem[Raychev et~al\mbox{.}(2015)]%
        {JSNice}
\bibfield{author}{\bibinfo{person}{Veselin Raychev}, \bibinfo{person}{Martin Vechev}, {and} \bibinfo{person}{Andreas Krause}.} \bibinfo{year}{2015}\natexlab{}.
\newblock \showarticletitle{Predicting Program Properties from "Big Code"}.
\newblock \bibinfo{journal}{\emph{SIGPLAN Not.}} \bibinfo{volume}{50}, \bibinfo{number}{1} (\bibinfo{year}{2015}).
\newblock
\showISSN{0362-1340}
\urldef\tempurl%
\url{https://doi.org/10.1145/2775051.2677009}
\showDOI{\tempurl}


\bibitem[{SALT Delft}(2021)]%
        {manytypes4py_github}
\bibfield{author}{\bibinfo{person}{{SALT Delft}}.} \bibinfo{year}{2021}\natexlab{}.
\newblock \bibinfo{title}{{Many-Types-4-Py Dataset}}.
\newblock \bibinfo{howpublished}{\url{https://github.com/saltudelft/many-types-4-py-dataset}}.
\newblock


\bibitem[{SALT Delft}(nd)]%
        {libsa4py}
\bibfield{author}{\bibinfo{person}{{SALT Delft}}.} \bibinfo{year}{n.d.}\natexlab{}.
\newblock \bibinfo{title}{libsa4py}.
\newblock \bibinfo{howpublished}{\url{https://github.com/saltudelft/libsa4py}}.
\newblock


\bibitem[Sapir and Contributors(2020)]%
        {chevrotainjava}
\bibfield{author}{\bibinfo{person}{Shahar Sapir} {and} \bibinfo{person}{The~Chevrotain Contributors}.} \bibinfo{year}{2020}\natexlab{}.
\newblock \bibinfo{title}{Chevrotain Java Grammar and CST Parser}.
\newblock \bibinfo{howpublished}{\url{https://www.npmjs.com/package/java-parser}}.
\newblock
\newblock
\shownote{Generates Concrete Syntax Trees for Java using Chevrotain}.


\bibitem[sitter Contributors(2018)]%
        {treesitter}
\bibfield{author}{\bibinfo{person}{Tree sitter Contributors}.} \bibinfo{year}{2018}\natexlab{}.
\newblock \bibinfo{title}{Tree-sitter: Incremental Parsing for Programming Tools}.
\newblock \bibinfo{howpublished}{\url{https://tree-sitter.github.io/tree-sitter/}}.
\newblock
\newblock
\shownote{Builds concrete syntax trees for many languages}.


\bibitem[{TIOBE}(2024)]%
        {TIOBE2024}
\bibfield{author}{\bibinfo{person}{{TIOBE}}.} \bibinfo{year}{2024}\natexlab{}.
\newblock \bibinfo{title}{TIOBE Index}.
\newblock \bibinfo{howpublished}{\url{https://www.tiobe.com/tiobe-index/}}.
\newblock


\bibitem[Wei et~al\mbox{.}(2020)]%
        {Lambdanet}
\bibfield{author}{\bibinfo{person}{Jiayi Wei}, \bibinfo{person}{Maruth Goyal}, \bibinfo{person}{Greg Durrett}, {and} \bibinfo{person}{Isil Dillig}.} \bibinfo{year}{2020}\natexlab{}.
\newblock \bibinfo{title}{LambdaNet: Probabilistic Type Inference using Graph Neural Networks}.
\newblock
\newblock
\showeprint[arxiv]{2005.02161}~[cs.PL]
\urldef\tempurl%
\url{https://arxiv.org/abs/2005.02161}
\showURL{%
\tempurl}


\bibitem[Wu et~al\mbox{.}(2024)]%
        {wu2024lemur}
\bibfield{author}{\bibinfo{person}{H. Wu}, \bibinfo{person}{C. Barrett}, {and} \bibinfo{person}{N. Narodytska}.} \bibinfo{year}{2024}\natexlab{}.
\newblock \showarticletitle{LEMUR: Integrating large language models in automated program verification}. In \bibinfo{booktitle}{\emph{ICLR}}.
\newblock
\urldef\tempurl%
\url{https://arxiv.org/abs/2310.06830}
\showURL{%
\tempurl}


\bibitem[Xu et~al\mbox{.}(2016)]%
        {XuPGM}
\bibfield{author}{\bibinfo{person}{Zhaogui Xu}, \bibinfo{person}{Xiangyu Zhang}, \bibinfo{person}{Lin Chen}, \bibinfo{person}{Kexin Pei}, {and} \bibinfo{person}{Baowen Xu}.} \bibinfo{year}{2016}\natexlab{}.
\newblock \showarticletitle{Python probabilistic type inference with natural language support}. In \bibinfo{booktitle}{\emph{Proceedings of the 2016 24th ACM SIGSOFT International Symposium on Foundations of Software Engineering}} (Seattle, WA, USA) \emph{(\bibinfo{series}{FSE 2016})}. \bibinfo{publisher}{Association for Computing Machinery}, \bibinfo{address}{New York, NY, USA}, \bibinfo{pages}{607–618}.
\newblock
\showISBNx{9781450342186}
\urldef\tempurl%
\url{https://doi.org/10.1145/2950290.2950343}
\showDOI{\tempurl}


\end{thebibliography}

\end{document}